\documentclass[preprint,aps,nofootinbib,floatfix,superscriptaddress,longbibliography]{revtex4-1}
\usepackage{amsmath,amssymb,amsfonts,mathrsfs,bbm,braket,xcolor,graphicx,graphics,latexsym,placeins,epsfig,multirow,makecell}
\usepackage{array}
\newcolumntype{P}[1]{>{\centering\arraybackslash}p{#1}}
\usepackage{footmisc}
\usepackage{comment}
\usepackage{bbold}
\usepackage{enumitem}
\setdescription{leftmargin=8pt}
\usepackage{bbold}
\usepackage{hyperref}

\usepackage{graphicx}
\usepackage{epsfig}
\usepackage{epstopdf}
\usepackage{amsfonts}
\usepackage{amssymb}
\usepackage{amsbsy}
\usepackage{amsmath}
\usepackage{mathrsfs}
\usepackage{latexsym}
\usepackage{bm}
\usepackage{color}
\usepackage{braket}
\usepackage{slashed}
\usepackage{pgfplots}
\usepackage{graphicx,subcaption}
\usepackage[normalem]{ulem}

\hypersetup{
	colorlinks=true,       
	linkcolor=blue,          
	citecolor=blue,        
	urlcolor=blue           
}
\DeclareMathOperator{\sech}{sech}
\usepackage{natbib}
\setcitestyle{square, comma, numbers, sort&compress}
\usepackage{float}
\usepackage{amsmath,amssymb}

\newcommand{\ba}{\begin{align}}
\newcommand{\ea}{\end{align}}

\def\3nab{\tilde{\nabla}}

\def\be {\begin{equation}}
\def\ee {\end{equation}}
\def\ba {\begin{eqnarray}}
\def\ea {\end{eqnarray}}

\newcommand{\sfr}[2]
{{\textstyle\frac{#1}{#2}}}

\newcommand{\barray}{\begin{array}}
\newcommand{\earray}{\end{array}}

\newcommand{\bea}{\begin{eqnarray}}
\newcommand{\eea}{\end{eqnarray}}

\begin{document}
\title{Analog charged black hole formation via percolation: Exploring cosmic censorship and Hoop conjecture}

\author{Nitesh Jaiswal}
\email{niteshphy@iitb.ac.in}

\author{S. Shankaranarayanan}
\email{shanki@iitb.ac.in}

\affiliation{Department of Physics, Indian Institute of Technology Bombay, Mumbai 400076, India}

\begin{abstract}
We investigate an analog model of charged black hole (BH) formation using the framework of classical percolation.  By analyzing the scaling behavior of key quantities, including surface gravity and Komar mass, we establish a robust correspondence between this analog system and gravitational collapse in general relativity. Our numerical simulations of the lattice model show excellent agreement with analytical predictions for the continuum limit, highlighting the potential of analog systems to capture essential features of BH physics.  Interestingly, we find that while geometric criteria related to the hoop conjecture are necessary, they are not sufficient for BH formation in our model.  Instead, the exponential growth of energy and cluster size emerges as the key indicator, suggesting a novel interpretation of the hoop conjecture and providing further support for cosmic censorship within our analog framework by ensuring horizon formation.  This work offers a fresh perspective on the organization of matter within BH event horizons and lays the groundwork for future quantum extensions that could shed light on Hawking radiation and the BH information paradox by linking entanglement entropy in quantum percolation models to BH entropy.
\end{abstract}

\maketitle

\section{Introduction}

Black holes (BHs) pose challenges to our understanding of fundamental physics~\cite{PhysRevLett.14.57,hawking1975particle,hawking1976breakdown}. 
The BH information paradox, stemming from the apparent conflict between quantum mechanical unitarity and the seemingly irreversible nature of BH evaporation~\cite{hawking1975particle,hawking1976breakdown}, highlights a critical gap in our knowledge, intertwined with our understanding of the nature of singularities~\cite{Senovilla:2014gza} and their fate in quantum gravity~\cite{preskill1992black,harlow2016jerusalem,Marolf:2017jkr,Ballik:2013uia,Chen:2014jwq,Christodoulou:2014yia,Bengtsson:2015zda,DAmbrosio:2020mut}. 
The cosmic censorship hypothesis further complicates this by suggesting singularities are always hidden by event horizons~\cite{Penrose:1969pc}. While quantum gravity is anticipated to provide a resolution~\cite{2022-Shanki.Joseph-GRG}, its development remains a major hurdle, prompting investigations into whether the paradox can be addressed without a complete theory of singularities~\cite{Ashtekar:2010qz,Ashtekar:2010hx,Okon:2014dpa,Mandal:2023kpu}.

Condensed matter systems provide a fertile ground for exploring this enigma. These systems exhibit phenomena remarkably analogous to BHs, including quasiparticle emergence, topological phase transitions, and complex entanglement dynamics~\cite{Schutzhold:2002rf, Unruh:2003ss, Unruh:2008zz, Sachdev:2010um, Sachdev:2015efa, Sachdev:2023try, Sachdev:2023fim, Shi:2021nkx, PhysRevX.7.031006, Pedernales, Viktor}. The emergence of quasiparticles, for instance, mirrors Hawking radiation~\cite{PhysRevLett.94.061302, PhysRevB.109.014309, Jacobson:1998he, tapo2019, tapo2022,Hawking:1982dh,Chamblin:1999tk, Chamblin:1999hg, tapo2015,Fairoos:2023hkk, Fan:2022bsq, Yerra:2022coh}. Studying information processing in these strongly correlated systems~\cite{Sachdev:1992fk,oppenheim2003thermodynamics, Saeed:2023ayw, PhysRevD.61.024017} can provide insights into information preservation and loss~\cite{1981-Unruh-PRL,2011-Barcelo.etal-LRR, Rosenberg:2020jde,PhysRevLett.110.101301, Stephens:1993an,Das:2008sy}. While these are analogs, not exact replicas, they serve as crucial experimental platforms for testing theoretical ideas~\cite{Lewenstein_2007,MunozdeNova:2018fxv,Steinhauer:2021fhb,Mertens:2022jij,Philbin,fischer23,fischer24, Yang:2024fql}.

Hawking radiation analogs in condensed matter differ from black holes in system size and the applicability of field-theoretic approximations. Furthermore, while the cosmic censorship hypothesis suggests that event horizons hide singularities in BHs, the absence of such horizons in analog systems raises questions about their ability to capture BH behavior fully. Therefore, assessing these analogs requires addressing key questions: When can a system mimic black hole geometry? How does a ground state evolve and reach a thermal steady state? What are the relevant timescales governing these processes?

In this work, we introduce a 1-D chiral quantum spin-chain model to explore key aspects of BH formation. The spin chain Hamiltonian, comprising $N$ spins with non-uniform interactions, features two competing terms: an inhomogeneous, short-range non-chiral term that sets the condition for BH-like formation and a homogeneous chirality operator with next-nearest-neighbor (NNN) interactions inducing long-range order. The interplay between these terms drives quantum phase transition (QPT)~\cite{2024-Wang.etal-AM}.

Chirality plays a crucial role in condensed matter physics, as exemplified by the chiral magnetic effect in Weyl semimetals~\cite{Hasan:2010xy,Bernevig-Book,Kharzeev:2013ffa}. In our model, the chiral term induces an imbalance in spin excitations, analogous to the charge imbalance in the chiral magnetic effect, which, in conjunction with the inhomogeneities of the non-chiral term, drives the system toward a topological QPT~\cite{2024-Wang.etal-AM}.  Chiral spin chains are also relevant for non-perturbative QFT via lattice techniques~\cite{Craig:2022cef,Davoudi:2022bnl}. Lattice chiral theories face fermion doubling issues~\cite{Nielsen:1981hk}, which our model circumvents using NNN interactions ($\chi_{j} = \vec{\sigma}_j \cdot \left(\vec{\sigma}_{j+1} \times \vec{\sigma}_{j+2}\right)$). This term acts as a long-range interaction, despite being constructed from local terms. This arises because the chirality operator effectively couples spins across multiple lattice sites, creating long-range correlations. The expectation value of the chirality operator, $\braket{\chi_j}$, encodes information about the system's global topology~\cite{PhysRevB.103.L060404, forbes2023exploring, Horner:2022sei}.  These long-range correlations bypass 
Nielsen-Ninomiya theorem~\cite{Nielsen:1981hk}.

The formation of BHs has been linked to critical phenomena, exhibiting power-law scaling near the threshold of BH formation, analogous to phase transitions with diverging correlation length~\cite{choptuik1993universality}. %
A key geometric criterion for predicting horizon formation in diverse scenarios, including non-symmetric collapse, is the \emph{hoop conjecture}~\cite{thorne2000gravitation}, which postulates that a mass ${\cal M}$ collapses to form a BH if and only if it can be enclosed by a circular hoop with a critical circumference ${\cal C}_{\rm critical} = 4 \pi {\cal M}$~\cite{flanagan1991hoop,hod2018status,bonnor1983hoop}, implying ${\cal C} \leq 4 \pi {\cal M}$. This suggests that sufficient mass concentration is crucial for horizon formation.

This work offers a novel perspective, connecting BH formation to percolation theory, a powerful framework for understanding cluster formation and connectivity at criticality, where large-scale connectivity emerges, modeling diverse physical phenomena~\cite{stauffer2018introduction}. 
We propose that gravitational collapse involves fluctuations leading to varying energy densities, and when both the density and connected clusters surpass a critical threshold, percolation occurs, forming an event horizon.
Specifically, we demonstrate a correspondence between the percolation threshold in a spin chain model~\cite{stauffer2018introduction, STAUFFER19791} and the emergence of an event horizon-like boundary, defined by NNN-dominated regions (interior) versus nearest-neighbor (NN)-dominated regions (exterior).  Within the NNN-dominated regime, we observe exponential growth in both interconnected spin clusters (analogous to matter aggregation) and the energy density within these clusters (akin to BH formation).  Furthermore, we find that the critical exponents associated with the total cluster energy and the percolation correlation length correspond to the BH mass and surface gravity exponents, respectively.  We discuss the implications for cosmic censorship and the hoop conjecture.

\section{Chiral spin chain}
\label{chiralchain}

The Hamiltonian of the 1-D chiral spin chain (with $\hbar = 1$) is:
{\small
\begin{eqnarray}
\label{eq:Hamiltonian01}
   \tilde{H} = - \sum_{j,k=1}^{N} \frac{U(j,k)}{2}\left(\sigma_j^x\sigma_k^x+\sigma_j^y\sigma_k^y\right) 
   + \sum_{j=1}^{N} \frac{v \, V(j,j+2)}{4} \chi_{j} \, ,  
\end{eqnarray}
}
where $\vec{\sigma}_j = (\sigma_j^x, \sigma_j^y, \sigma_j^z)$ denotes the Pauli matrix vector operator of the $j$-th site with periodic boundary conditions (PBC) ($\vec{\sigma}_j=\vec{\sigma}_{j+N}$), 
$\chi_{j}=\vec{\sigma}_j \cdot \left(\vec{\sigma}_{j+1} \times \vec{\sigma}_{j+2}\right)$ is the chirality operator~\cite{Horner:2022sei, forbes2023exploring, tsomokos2008chiral, d2005chiral, yang2024smartholesanalogueblack, Daniel2024boy}, $v$ is a \emph{dimensionless constant}, $U(j,k)$ and $V(j,j+2)$ are site-dependent coupling constants that introduce inhomogeneity. The first term in the RHS describes an inhomogeneous spin-1/2 XX interaction between sites $j$ and $k$, while the second term represents NNN three-spin chiral interaction involving spins at sites $j$, $j+1$, and $j+2$~\cite{Horner:2022sei}. As we show, the physical properties of the system are determined by the relative strength of the two coupling constants $U(j,k)$ and $V(j,j+2)$. We now rewrite the above Hamiltonian: 
\begin{align}
\label{eq:Hamiltonian02}
  \tilde{H} &= \sum_{j,k=1}^{N} {V(j,j+2)}  H(j,k)\, , \\
 H(j,k) & = - {u (j, k)} \left(\sigma_j^x\sigma_k^x+\sigma_j^y\sigma_k^y\right)/2 + {v} \, \chi_{j}/4  \, ,
\end{align}
where $u(j,k)= {U(j,k)}/{V(j,j+2)}$ is dimensionless. Since the system is 1-D, the coupling depends only on the distance between the sites, i.e., $u(j,k)\equiv u(|j-k|)$. If $V(j,j+2)$ is well-defined for all $j$, the key properties of the system are governed by the reduced Hamiltonian $H(j,k)$ while $V(j,j+2)$ scales the energy of the system. As shown in Appendix \ref{EBHappendix}, $V(j,j+2)$ is a conformal factor in the continuum limit to the spatial part of the formed BH metric~\cite{Chandran:2020gcd}. Hence, Hamiltonian \eqref{eq:Hamiltonian01} reduces to the \emph{dimensionless Hamiltonian}: 
\begin{equation}
  H=\sum_{j,k=1}^{N}   H(j,k) \, ,
  \label{Hamiltonian} 
\end{equation}
where $j$, $k$ are now \emph{identified as NN sites} and the coupling $u(|j-k|)$ makes the system inhomogeneous. The Hamiltonian \eqref{Hamiltonian}, while inhomogeneous, exhibits translational invariance since $u(|j-k|)$ depends solely on the distance between sites. A Jordan-Wigner transformation~\cite{Jordan:1928wi, shigu} yields a Hamiltonian with four-fermion interaction. Applying mean field theory (details are contained in Appendix \ref{JWappendix} and \ref{EBHappendix}), leads to: 
%
\begin{equation}
\!\!\! H_{MF}=- \sum_{j=1}^{N}\left[u(j,j+1)c^\dagger_jc_{j+1} + \frac{iv}{2}c^\dagger_jc_{j+2}\right]+H.C.,\label{Mfhamiltcn}
\end{equation}
where $c^\dagger_j$ and $c_j$ are the fermionic creation and annihilation operators satisfying the standard anticommutation relations. 
Choosing $u$ consistent with PBC, a discrete Fourier transform diagonalizes this Hamiltonian:
\begin{equation}
     H_{MF}=\sum_{p}E(p)c^\dagger_p c_p~,~E(p)=\tilde{u}(p)+\tilde{u}^{\ast} (p) +v \sin p~.\label{Mfhamilt}
\end{equation}
Here $p\in [-\pi, \pi)$, $E(p)$ is the dispersion relation, and $\tilde{u}(p)= \mathcal{U}e^{ip}$ is the Fourier transform of $u(|(j+1)-j|)=\mathcal{U}$. The Fermi points ($E(p) = 0$) are typically at $p_{R, L}=\pm \pi/2$.  However, for  $|v|>|\mathcal{U}|$, two additional crossings occur at $p_1=\sin^{-1}\left(\frac{\mathcal{U}}{v}\right)$ and $p_2=\pi - p_1$.  The ground state energy density in the thermodynamic limit is:
\begin{equation}
    \rho_0 = \frac{1}{2\pi}\int_{p:E(p)<0} dp E(p)\quad = \,\Biggl\{
  \begin{array}{@{}ll@{}}
    \frac{2\mathcal{U}}{\pi}, & v\leq \mathcal{U} \\
    -\frac{\mathcal{U}^2 + v^2}{v \pi}, & v>\mathcal{U}
  \end{array}~.
\end{equation}
A discontinuity in the second derivative of $\rho_0$ at $|v|=|\mathcal{U}|$ signals a second-order QPT~\cite{2024-Wang.etal-AM}. 

\section{Analog BH}

The model admits a geometric interpretation in the context of BH physics. To establish this connection, we partitioned the lattice sites of $H_{MF}$ in (\ref{Mfhamiltcn}) into alternating sublattices A and B, a procedure that facilitates the emergence of Dirac-like quasiparticles upon taking the continuum limit. This limit is achieved by expanding the mean-field Hamiltonian using a Taylor series around the Fermi points, where the characteristic Dirac cone structure arises. In the low-energy regime, the linear component of the dispersion relation dominates, while higher-order terms are negligible. This approximation yields an effective Dirac Hamiltonian defined on a 2-D spacetime with the following metric (details in Appendix \ref{EBHappendix}):
\begin{equation}
\label{eq:AnalogBHmetric01}
    ds^2 = g(x) dt^2- 2 \sqrt{1 - g(x)} \frac{dt \, dx}{\mathcal{U}(x)} -\frac{dx^2}{\mathcal{U}^2(x)}~,
\end{equation}
where $(t,x)$ are dimensionless coordinates, $g(x) = 1- {v^2}/{\mathcal{U}^2(x)}$. This is the Gullstrand-Painl\'eve coordinates of spherically symmetric space-time~\cite{Shankaranarayanan:2003ya}. Employing the coordinate transformation $(t,x)\rightarrow$ $(\tau,x)\rightarrow$ $(\tau,X)$ via
\begin{equation}
    \tau(t,x)=t-\int_{x_0}^x \frac{dz}{\mathcal{U}^2(z)} \frac{v}{g(z)}~,~dX=\frac{1}{g(x)} \frac{dx}{\mathcal{U}(x)}~,
\end{equation}
brings the metric to the following two generic forms:
\begin{equation}
 \label{metrictau}
 ds^2 = g(x)d\tau^2 - d\mathcal{X}^2/g(x) = g(x)\left(d\tau^2 - dX^2\right),
\end{equation}
where $d\mathcal{X} = {dx}/{\mathcal{U}(x)}$.
Until this point, ${\mathcal{U}(x)}$ is arbitrary, however, to establish a meaningful correspondence with an analog BH solution, we need to make a specific choice of ${\mathcal{U}(x)}$. In the rest of this work, we choose the form of $\mathcal{U}(x)$ that is smooth, and asymptotically decays:
\begin{equation}
\label{eq:formU}
\mathcal{U}(x)= \sech\left({x}/{\sigma}\right)/{\sqrt{2\sigma}}\, ,
\end{equation}
where $\sigma$ is a constant.  For $\sigma< \sigma_c \equiv {1}/{(2v^2)}$, the  $\tau-\tau$ component of the metric (\ref{metrictau}) vanishes at two distinct points. This condition signifies the presence of two horizons, defined by $g(x)=0$ and located at:
\begin{equation}
    x_{c\pm} = \pm \sigma\cosh^{-1}\left(\sqrt{\sigma_c/\sigma}\right), \quad \text{if} \quad \sigma<\sigma_c~.
\end{equation}
The emergence of two horizons exhibits a striking resemblance to the Reissner-Nordstr\"om (RN) BH solution in 4-D spacetime. The RN metric is given by~\cite{reissner1916,nordstrom1918}:
\begin{equation}
d\tilde{s}^2 = f(r) \left(-dt^2+dr_{\ast}^2\right)+r^2 d\Omega_2^2\, ,
\end{equation}
where $f(r) = 1- {2M}/{r}+ {Q^2}/{r^2}, r_{\ast}=\int {dr}/{f(r)}$ and, the two horizons are located at $r_{\pm}=M\pm\sqrt{M^2-Q^2}$.

This is the first key result of this work, which we want to discuss in detail: First, the continuous chiral analog model mimics the 4-D RN BH. At $\sigma = \sigma_{\rm c}$, the two horizons coincide, similar to the extremal RN BH where $r_{\pm}=2M$. At this critical value, we find $x_{c\pm}=0$, mirroring the behavior of the extreme RN solution. In addition, for $\sigma>\sigma_{\rm c}$, the spacetime remains regular, exhibiting no horizons or naked singularity. Thus, $\sigma$ effectively plays a role analogous to the charge-to-mass ratio $(Q/M = \sigma/\sigma_c)$ in the RN solution. This analogy highlights the crucial role of $\sigma$ in determining the spacetime structure, particularly the presence or absence of horizons, similar to how $M$ and $Q$ govern the RN BH. This avoidance of naked singularities for $\sigma > \sigma_c$ is consistent with cosmic censorship.  In this analog model, $\sigma$ effectively enforces a form of cosmic censorship, ensuring that when \emph{the charge} represented by $\sigma$ exceeds a critical value relative to \emph{the mass} (related to $\sigma_c$), no naked singularity develops.

Furthermore, despite the global neutrality of the chiral spin chain model \eqref{Hamiltonian}, effective local charges can arise at each site. This localization stems from the inhomogeneous coupling between sites, resulting in a non-uniform charge distribution across the system. Furthermore, the non-uniformity generates an effective electric field with a $\sech(x)$ profile, exhibiting a characteristic decrease in strength with distance. This behavior resembles the screening effect in electrostatics, where fields from other charges partially or fully cancel the electric field from a given charge.  The non-homogeneous coupling reduces the effective electric field, providing valuable insights into how local interactions influence the field distribution within the spin chain. 

To further understand the connection between the analog model and RN BH, we evaluate \emph{three geometric quantities} for the analog metric \eqref{eq:AnalogBHmetric01}: the Kretschmann scalar ($\mathcal{K}$), surface gravity ($\kappa_g$), and Komar mass $(M_{K}$)~\cite{Wald:1984rg}. $\mathcal{K}$ is a quadratic scalar invariant of the Riemann tensor~\cite{Wald:1984rg,carroll2019spacetime}, defined as $\mathcal{K}=R_{\mu\nu\rho\eta}R^{\mu\nu\rho\eta}$. For metric \eqref{eq:AnalogBHmetric01}, $\mathcal{K}= 1/(\sigma^4 \sigma_c^2)$. Depending on $v$, $0 < \mathcal{K} < \infty$. 
The line-element \eqref{eq:AnalogBHmetric01} has a timelike Killing vector $K^\mu$, $\kappa_g$ at event horizon is defined as $\kappa_g^2 = -\frac{1}{2}(\nabla_\mu K_\nu)(\nabla^\mu K^\nu)$:
%
\begin{equation}
    \label{surfg}
\kappa_g^2 = \frac{1}{8 \sigma  \sigma_c^2} \left[6
\sinh ^2\left(\frac{x}{\sigma }\right) - 
  \frac{1}{\sigma} \tanh ^2\left(\frac{x}{\sigma }\right) g^2(x) \right] ~ .
\end{equation}
The Komar mass is~\cite{carroll2019spacetime},
\begin{equation}
    M_{K}=\int_S dx \sqrt{|g|} n_\mu \nabla_\nu(-\nabla^\nu K^\mu)~,
\end{equation}
where $S$ is a constant time slice, $g$ is the determinant of the metric, and $n^\mu$ is a unit vector orthogonal to $S$. The boundary of $S$ has two contributions, one from an inner boundary of the horizon which is BH mass evaluated as
\begin{equation}
\!\!\! M_{BH}=\frac{1}{\sigma_c} \sqrt{\frac{2}{\sigma }}  \tan^{-1}\left[\tanh \left(\frac{1}{2} \cosh^{-1}\left[\sqrt{\frac{\sigma_c}{{\sigma }}} \right]\right)\right],
\label{KM}    
\end{equation}
and the second from an outer boundary at infinity. Later, these quantities will be employed to investigate the nature of the phase transition in the spin-chain and determine whether it exhibits characteristics analogous to a RN BH.
In Appendix \ref{EBHappendix}, we discuss limits when $v$ or ${\cal U}$ vanish or diverge, leading to Carrollian and shockwave limits.

Having established that the chiral spin chain indeed mimics RN BH, the immediate question that needs to be answered is: Can such a system be \emph{observed in the laboratory}? Primarily, where the generation of Hawking radiation often relies on probabilistic processes, a significant challenge arises from the need for active feedback mechanisms. These mechanisms dynamically route quantum systems based on the outcomes of previous interactions, akin to \emph{active switching} in quantum computing~\cite{Kieling2009}. This is particularly demanding in linear wave phenomena scenarios, which are common in analog gravity experiments.

In the rest of this work, we use
percolation theory to address this challenge. By carefully designing the initial configuration of the system and employing interactions, we demonstrate the potential to eliminate the need for active feedback. This is achieved by leveraging a static, pre-determined network structure within the analog system. 
These regions then interact according to the lattice geometry. By carefully choosing this geometry, we can ensure that a percolating cluster emerges, enabling the desired behavior to arise spontaneously.

\section{Percolation model of a chiral spin chain}

We now construct a classical bond percolation model~\cite{roy2019exact,roy2019percolation} that acts as a proxy for the tight-binding model (\ref{Hamiltonian}). Specifically, like in Refs. \cite{welsh2018simple, logan2019many}, we interpret the Hamiltonian (\ref{Hamiltonian}) in the Fock space as a tight-binding model. A classical percolation model is typically used to study the behavior of connected clusters in a random medium. It involves a lattice where each site or bond is randomly occupied with a probability $\mathcal{P}$. A large connected cluster emerges at a critical probability $\mathcal{P}_c$, marking a phase transition.

\begin{figure}
     \centering
     \begin{subfigure}[b]{0.2\textwidth}
         \centering
         \includegraphics[width=\textwidth]{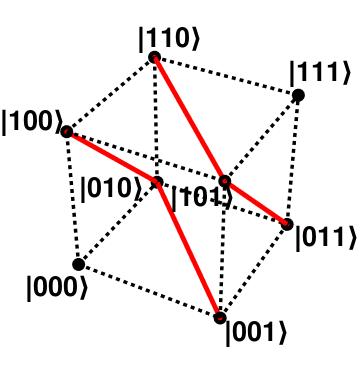}
     \end{subfigure}
  \hspace*{0.5cm}
     \begin{subfigure}[b]{0.2\textwidth}
         \centering
         \includegraphics[width=\textwidth]{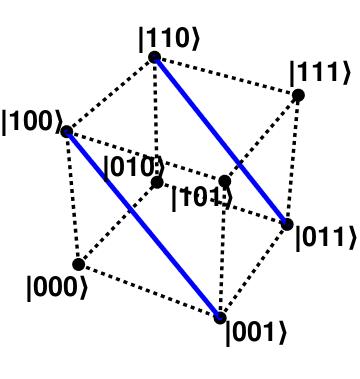}
     \end{subfigure}
     \caption{Fock-space cube illustrating system behavior for dominant NN interactions (red edges in the left figure) and NNN interactions (blue edges in the right figure) in a three-spin system. Nodes represent basis states, and active edges represent allowed transitions.}
     \label{activedges}
\end{figure}

We define a graph in Fock space to apply this to our system (\ref{Hamiltonian}). Each node in this graph represents a basis state of the quantum system, specifically a Fock state of fermions with occupation numbers $n_j=0, 1$. The edges between these nodes correspond to non-zero matrix elements of the Hamiltonian. In other words, an edge connects two nodes if the Hamiltonian has a non-zero transition amplitude between the corresponding Fock states. The Hamiltonian (\ref{Hamiltonian}) is thus represented on this Fock-space graph by the weights associated with the edges:
\begin{equation}
    H_{FS}=\sum_{I\neq K} T_{IK} \ket{I}\bra{K}~,  \label{percolationhamilt}
\end{equation}
where $H_{FS}$ in Fock space acts on a graph where nodes represent Fermionic basis states $\ket{I} = \ket{n_1,n_2,\dots,n_N}$. Edges in this graph correspond to off-diagonal matrix elements $T_{IK}$ of $H_{FS}$, representing hopping between states $\ket{I}$ and $\ket{K}$, i.e., $T_{IK} = \langle I | H_{\rm MF} | K \rangle$, 
%
where the NN and NNN fermionic creation and annihilation operator terms inside the bracket correspond to the mean-field Hamiltonian (\ref{Mfhamiltcn}). 
Active edges on the graph represent either NN interactions ($\mathcal{U}$ dominant) or NNN interactions ($v$ dominant).  When $\mathcal{U}$ dominates, only edges between NN nodes are active. When $v$ dominates, diagonal edges representing NNN interactions become active. Fig. \ref{activedges} illustrate these active edges for $N=3$, with red (left) and blue (right) lines indicating NN and NNN interactions, respectively, and the cube nodes represent Fock states. The percolation cluster is the largest set of connected nodes. A phase transition occurs when the NN and NNN interaction strengths become equal, marking the onset of a \emph{percolation transition}~\cite{Schrenk, Prakash}.

\subsection{Percolation exponents}
\label{percoexpo}
\begin{figure}
\centering
\begin{subfigure}[b]{0.22\textwidth}
         \centering
         \includegraphics[width=\textwidth]{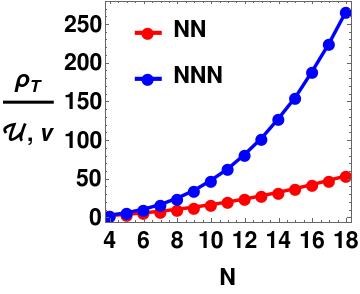}
     \end{subfigure}
  \hspace*{0.5cm}   
\begin{subfigure}[b]{0.22\textwidth}
         \centering
         \includegraphics[width=\textwidth]{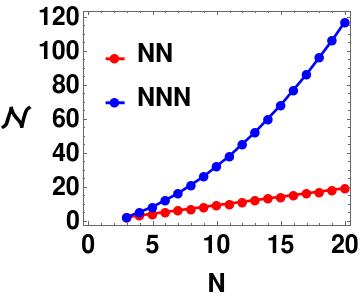}
     \end{subfigure}
    \caption{Normalized cluster energy density (left) and ${\cal N}$ (right) vs. system size N. Red (blue) circles: Dominant NN (NNN) interactions.}
    \label{totalcluster}
\end{figure}

To analyze cluster growth, we examine total energy density $(\rho_T)$ and number of clusters $({\cal N})$ as a function of system size $N$. To keep the discussion transparent, we consider a deterministic percolation scenario where bond activation is governed by interaction strengths, with no randomness in forming connections. 
Therefore, we can extract critical exponents by systematically increasing $N$. To go about this we need to quantify how densely the energy is concentrated within a cluster including the interaction energy (IE). The energy density of an individual cluster is defined as:
\begin{equation}
\!\!\!\! \text{Energy density} = \frac{\text{\# of connected bonds}}{\text{\# of connected states}} \times\,\text{IE},
\end{equation}
where ``\# of connected states" refers to the distinct states within a given cluster that are connected via one or more bonds and IE is $\mathcal{U}$ for NN and $v$ for NNN interactions.  $\rho_T$ is obtained by summing the energy densities of all individual clusters. As shown in Fig. \ref{totalcluster}, we see two distinct features: When NN interactions dominate, the normalized total energy density of the clusters $\rho_T/\mathcal{U}$, shown in red (left) and ${\cal N}$ shown in red (right) increases linearly with $N$, indicating a simple connectivity structure. However, when 
NNN interactions dominate, the normalized total energy density of the clusters  $\rho_T/v$ shown in blue (left) and the number of clusters shown in blue (right) exhibit exponential growth, characteristic of a BH-like phase with intricate connectivity. The contrasting growth provides strong evidence for BH phase interpretation.

\begin{figure}
     \centering
     \begin{subfigure}[b]{0.21\textwidth}
         \centering
         \includegraphics[width=\textwidth]{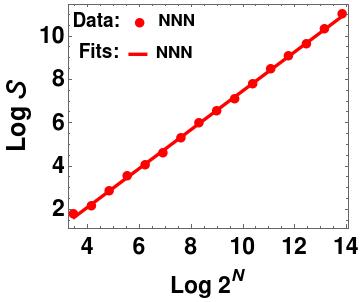}
     \end{subfigure}
       \hspace*{0.5cm}
     \begin{subfigure}[b]{0.22\textwidth}
         \centering
         \includegraphics[width=\textwidth]{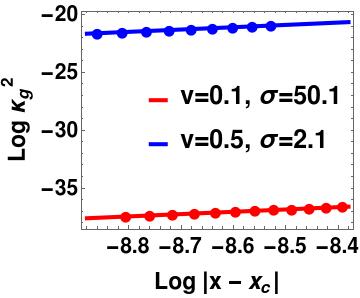}
     \end{subfigure}
     \caption{Plot of $\log \mathcal{S}$ vs $\log 2^N$ (left) and $\log \kappa_g^2$ vs $\log |x-x_c|$ (right). Solid lines are fits to numerical (left) and analytical values using Eq. (\ref{surfg}) (right). }
     \label{CorrelateKretsch}
\end{figure}
\begin{figure}
\centering  
     \begin{subfigure}[b]{0.22\textwidth}
         \centering
         \includegraphics[width=\textwidth]{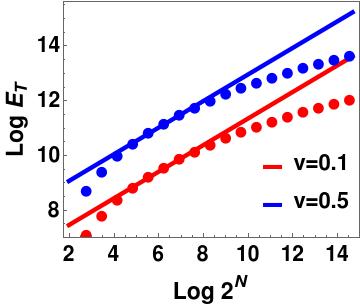}
     \end{subfigure}
  \hspace*{0.15cm}
     \begin{subfigure}[b]{0.238\textwidth}
         \centering
         \includegraphics[width=\textwidth]{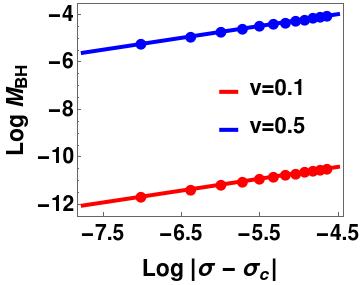}
     \end{subfigure}
     \caption{Plot of $\log E_T$ vs $\log 2^N$ (left) and $\log M_{BH}$ vs $\log |\sigma - \sigma_c|$ (right). Solid lines are the fits to the numerical (left) and analytical values using Eq. (\ref{KM}) ({right}).}
     \label{EnergyKomar}
\end{figure}

To further understand the relation between the phase transition of the analog model and RN BH, we analyze the correlation length critical exponent ($\nu_P$) which characterizes the behavior of the correlation length $\xi$ near the percolation transition. Here, $\xi$ corresponds to the size of the largest cluster ($\mathcal{S}$). Near the critical point, $\xi$ follows power law: $\xi \sim |\mathcal{P}-\mathcal{P}_{c}|^{\nu_P}$~\cite{Schrenk, Prakash}. Since at critical point the system is scale-invariant, implying that $\xi$ must be comparable to the number of Fock states, hence~\cite{Schrenk, Prakash}:
\begin{equation}
	\xi \sim |\mathcal{P}-\mathcal{P}_{c}|^{\nu_P} \sim 2^{N}~.
\end{equation}
By plotting $\log \mathcal{S}$ versus $\log(2^N)$ in Fig. \ref{CorrelateKretsch} (left), we determine $\nu_P$ from the slope of the linear fit.  For the dominant NNN interaction, we find $\nu_P = 1.11$. This exponent governs the divergence of the correlation length at the critical point and will be used to determine critical exponent corresponding to the total energy ($E_T$) of the system. 
%
%
The total energy, $E_T$, is calculated by multiplying $\rho_T$ with the total number of states. Since, $E_T$ changes due to IE, like $\xi$, $E_T$ also diverges at the critical point allowing us to write $E_{T} \sim |\mathcal{P}-\mathcal{P}_{c}|^\beta \sim (2^{N})^{\frac{\beta}{\nu_P}}$. The left Fig. \ref{EnergyKomar} is the plot of
$\log E_T$ versus $\log(2^N)$ for two values of $v$, and $\mathcal{U}$ given by~\eqref{eq:formU}. From the slope of the linear fit around the transition, we obtain $\beta/\nu_P$ yielding $\beta = 0.53$ for the dominant NNN interaction case.

\begin{figure}
     \centering
     \begin{subfigure}[b]{0.2\textwidth}
         \centering
         \includegraphics[width=\textwidth]{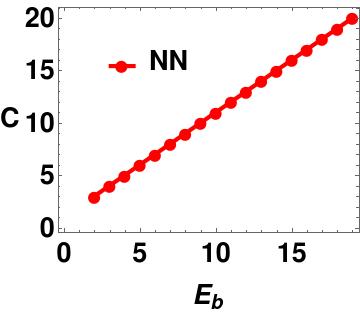}
     \end{subfigure}
       \hspace*{0.5cm}
     \begin{subfigure}[b]{0.2\textwidth}
         \centering
         \includegraphics[width=\textwidth]{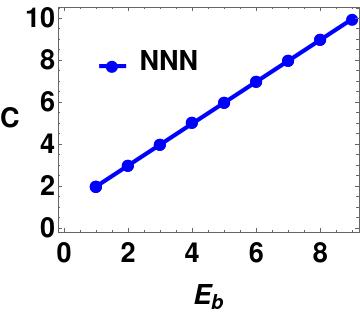}
     \end{subfigure}
     \caption{Plot of circumference $C$ of the smallest cluster and the energy of connected bonds $E_b$ for dominating NN (left) and NNN (right) phase. }
     \label{Hoopfig}
\end{figure}

Fig. \ref{Hoopfig} shows the number of connected bonds $E_b$, and the number of states $C$ in the smallest cluster. 
The equal slopes of the plotted lines for both the NN- and NNN-dominated phases suggest that the hoop conjecture~\cite{thorne2000gravitation,flanagan1991hoop,hod2018status,bonnor1983hoop}, should be satisfied, requiring $\mathcal{U}\geq 1$ in the NN phase and $v\geq 1$ in the NNN phase. However, our analysis reveals that $v\geq \mathcal{U}$, a condition necessary for BH formation, is \emph{not} satisfied. This indicates that the hoop conjecture, within our percolation model, does not impose stringent conditions favoring one phase over the other for BH formation.  While geometrically satisfying the hoop conjecture within the percolation model appears to be a \emph{necessary} condition for the emergence of a BH-like object, our results suggest it is not \emph{sufficient}~\cite{Hod:2020eau}. 

\begin{table}[ht!]
\centering
\begin{tabular}{ |c|c||c|c|  }
 \hline
 \multicolumn{2}{|c||}{\rule{0pt}{4mm}\textbf{Percolation Model (NNN phase)}\rule[-2mm]{0pt}{2mm}} & \multicolumn{2}{|c|}{\textbf{Gravitational Collapse (BH phase)}}\\
 \hline
\rule{0pt}{4mm}\textbf{Quantity} & \textbf{Critical Exponent}\rule[-2mm]{0pt}{2mm}  & \textbf{Quantity} & \textbf{Critical Exponent}\\
 \hline
 \makecell{Correlation length ($\xi$): \vspace{-2mm}\\ $\xi \sim |\mathcal{P}-\mathcal{P}_{c}|^{\nu_P}$ } & $\nu_P=1.11$  & \makecell{Surface gravity ($\kappa_g$): \vspace{-2mm}\\ $\kappa_g \propto (x - x_c)^{\gamma}$} &   $\gamma=1.00$ \\
 \hline
 \makecell{Total energy ($E_T$): \vspace{-2mm}\\ $E_T \sim |\mathcal{P}-\mathcal{P}_{c}|^{\beta}$ } & $\beta=0.53$  & \makecell{Black hole mass ($M_{BH}$): \vspace{-2mm}\\ $M_{BH} \propto (\sigma - \sigma_c)^{\gamma_M}$} &   $\gamma_M=0.5$ \\
 \hline
\end{tabular}
\caption{Comparison of critical exponents between the percolation model and gravitational collapse.}
\label{tab:critical_exponents}
\end{table}

A key finding of this work is the exceptional agreement between the numerically calculated scaling exponents for the lattice model (with $N \leq 20$) and the corresponding analytical results for the continuum limit.  This agreement is clearly demonstrated by the comparison of surface gravity (Fig. \ref{CorrelateKretsch}, right) and Komar mass (Fig. \ref{EnergyKomar}, right), and is further detailed in Table \ref{tab:critical_exponents}. As $N$ increases, this comparison is expected to become exact.

\section{Conclusions}

Our results offer significant insights into the analog model and its connection to BH physics. The exceptional agreement between numerical results for finite lattice sizes and analytical predictions for the continuum model demonstrates the robustness of this analog system in capturing key features of gravitational collapse. This is consistent with established mappings between quantum spin chains and classical percolation models~\cite{roy2019exact,roy2019percolation,welsh2018simple}, where a phase transition at the percolation threshold mirrors critical phenomena in the spin chain. This transition reflects a change in lattice connectivity analogous to the behavior observed in quantum systems at criticality.

Classical percolation effectively models how local connectivity rules generate macroscopic structures and critical behavior near transitions. This directly parallels black hole formation, accurately reproducing critical exponents. The percolating cluster emergence mirrors event horizon formation, where increasing interaction strength analogously represents growing gravitational influence.
The close alignment of scaling exponents derived from the analog model with those of surface gravity and Komar mass in the BH context further validate this framework. Crucially, the exponential cluster growth observed in analogous to matter aggregation during stellar collapse.

Our results also shed light on the hoop conjecture \cite{thorne2000gravitation,bonnor1983hoop,flanagan1991hoop,hod2018status} and opens a new avenue for exploring BH-like phenomena in complex systems by highlighting the importance of additional criteria beyond the geometric hoop conjecture. Specifically, the exponential growth of clusters and energy is a key indicator of BH formation in this analog system, potentially acting as a proxy for the \emph{hoop} itself. Furthermore, the observed formation of a horizon-like structure, coupled with the absence of naked singularities for $\sigma > \sigma_c$, provides further support for cosmic censorship within this analog framework.

This analysis provides a step towards understanding matter/energy organization within a BH event horizon, a region inaccessible to direct observation in astrophysical BHs. By simulating BHs in the lab via this analog system, we can begin to probe these questions without needing a full theory of BH singularities. This work focuses on the classical aspects of the percolation and gravity sides. Future quantum extensions, incorporating superposition and entanglement, will allow us to explore Hawking radiation and BH information paradox, especially at the critical value $\sigma =  \sigma_c$ when the horizons degenerate and surface gravity vanishes. Specifically, the entanglement entropy arising from quantum correlations between sites within and outside forming clusters may correspond to BH entropy~\cite{Chandran:2020gcd}, offering insights into the information encoded within the horizon. These quantum extensions are under investigation.

\section*{Acknowledgments}

The authors thank G. Baskaran, Indranil Chakraborty, K. Hari, P. George Christopher, Susobhan Mandal, Tausif Parvez, Tapobrata Sarkar, and Semin Xavier for discussions and comments on the earlier draft.  We also thank the authors of Ref.~\cite{forbes2023exploring} for email correspondence. The work is supported by SERB-CRG (RD/0122-SERB000-044).


\appendix

\section{Jordan-Wigner and Fourier transformations}
\label{JWappendix}

Our starting point is a system of $N$ spin-1/2 particles arranged in a one-dimensional periodic chain with inhomogeneous nearest-neighbor (NN) and chiral interactions. The Hamiltonian governing the system is given by
\begin{equation}
    \tilde{H}=-\sum_{j,k=1}^{N}\frac{U(j,k)}{2}\left(\sigma_j^x\sigma_k^x+\sigma_j^y\sigma_k^y\right)+\sum_{j=1}^{N-2}\frac{v V(j,j+2)}{4} \left(\vec{\sigma}_j \cdot \left(\vec{\sigma}_{j+1} \times \vec{\sigma}_{j+2}\right)\right)~,
\end{equation}
where $v$ is a constant, $U(j,k)$ and $V(j,j+2)$ are site-dependent coupling constants introducing spatial inhomogeneity into the system. Here, the operators $\sigma_j^{x,y,z}$ represent the Pauli matrices at the $j$-th site, and $\vec{\sigma}_j = (\sigma_j^x, \sigma_j^y, \sigma_j^z)$ denotes the Pauli matrix vector of the $j$-th spin. The first term of the Hamiltonian describes an inhomogeneous spin-1/2 XX interaction between sites $j$ and $k$, while the second term represents a three-spin chiral interaction involving consecutive spins at sites $j$, $j+1$, and $j+2$. The physical properties of the system are determined by the relative strength of the two coupling constants $U(j,k)$ and $V(j,j+2)$. 
To simplify the analysis, we can rescale the Hamiltonian by the coupling constant $V(j,j+2)$ which yields 
\begin{equation}
    \tilde{H} = \sum_{j,k=1}^{N} V(j,j+2) H(j,k)~,\quad \text{with}\quad  
 H(j,k)  = - \frac{u (j, k)}{2} \left(\sigma_j^x\sigma_k^x+\sigma_j^y\sigma_k^y\right)+ \frac{v}{4}  \left(\vec{\sigma}_j \cdot \left(\vec{\sigma}_{j+1} \times \vec{\sigma}_{j+2}\right)\right)~, \label{Hamilt}
\end{equation}
where $u(j,k)=\frac{U(j,k)}{V(j,j+2)}$. Since we have a one-dimensional system, without loss of generality, we can restrict the coupling constants $u(j,k)\equiv u(|j-k|)$, and $V(j,j+2)\equiv V(|(j+2)-j|)$, to depend only on the distance between the sites so that the system remains translationally invariant. Later in this section while performing the Fourier transformation, we will show that $V(j,j+2)$ only modifies the overall energy scale of the system. In the continuum limit, the system is described by a massless, minimally coupled Dirac field in a (1+1)-dimensional curved spacetime. Since the curved spacetime is locally conformal to the Minkowski space, the dynamics of the Dirac field remain unchanged up to a simple rescaling. Thus, the coupling $V(j,j+2)$ merely acts as a conformal factor to the spatial metric in the continuum limit. The key properties of the Hamiltonian $\tilde{H}$ remain unchanged if we remove this conformal factor. Hence, we can work with the rescaled Hamiltonian of the form
\begin{equation}
   H=-\sum_{j,k=1}^{N}\frac{u(|j-k|)}{2}\left(\sigma_j^x\sigma_k^x+\sigma_j^y\sigma_k^y\right)+\sum_{j=1}^{N-2}\frac{v}{4} \left(\vec{\sigma}_j \cdot \left(\vec{\sigma}_{j+1} \times \vec{\sigma}_{j+2}\right)\right)~,\label{redHamilt} 
\end{equation}
which preserves the essential physics of the system while eliminating unnecessary rescaling. Here, $j$, $k$ are now identified as NN sites and the coupling $u(|j-k|)$ makes the system inhomogeneous and becomes the source of curvature in the continuum limit. In order to discuss the mean-field theory and analytic results, we apply a Jordan-Wigner transformation which maps spin-1/2 Pauli operators to spinless fermions defined as
\begin{equation}
    \sigma_{j}^{+}=\exp{\left(-i\pi\sum_{l=1}^{j-1}c_{l}^{\dagger}c_{l}\right)}c_{j}^{\dagger}~,\quad \sigma_{j}^{-}=\exp{\left(i\pi\sum_{l=1}^{j-1}c_{l}^{\dagger}c_{l}\right)}c_{j}~,\quad  \sigma_{j}^{z}=1-2c_{j}^{\dagger}c_{j}~,
\end{equation}
where $\sigma_{j}^{\pm}= (\sigma_j^x\pm \sigma_j^y)/2$ and $c_j$ are fermionic operator satisfying anti-commutation relations $\{c_j,c_k^\dagger\}=\delta_{jk}$ and $\{c_j,c_k\}=\{c_j^\dagger,c_k^\dagger\}=0$. The mapping of the chirality operator, \(\chi_j = \vec{\sigma}_j \cdot \left(\vec{\sigma}_{j+1} \times \vec{\sigma}_{j+2}\right)\), to the Fermionic Hamiltonian has been extensively discussed in \cite{Horner:2022sei,forbes2023exploring}. However, since the first term in the original Hamiltonian is non-trivial, we focus here on its explicit mapping to the fermionic representation. For completeness, we derive the mapping for both the original Hamiltonian and the rescaled Hamiltonian. The first terms of Eq. (\ref{Hamilt}) and the rescaled Hamiltonian in Eq. (\ref{redHamilt}) are given by:
%
\begin{equation}
    \tilde{H}_{XX} = -\sum_{j,k=1}^{N}V(|(j+2)-j|)u(|j-k|)\left(c_{j}^{\dagger}c_{k}+c_{k}^{\dagger}c_{j}\right)~,\quad  H_{XX} = -\sum_{j,k=1}^{N}u(|j-k|)\left(c_{j}^{\dagger}c_{k}+c_{k}^{\dagger}c_{j}\right)~.\label{xxhamilt}
\end{equation}
Although the model is inhomogeneous, the above Hamiltonians in Eq. (\ref{xxhamilt}) remain translationally invariant because $u(|j-k|)$ depend solely on the distance between sites. Consequently, it can be exactly diagonalized using the discrete Fourier transformation
\begin{equation}
    c_j = \frac{1}{\sqrt{N}}\sum_{p\in [-\pi,\pi)} e^{ipj}c_p~,\quad c_j^\dagger = \frac{1}{\sqrt{N}}\sum_{p\in [-\pi,\pi)} e^{-ipj}c_p^\dagger~,
\end{equation}
where the quasi-momentum $p$ takes the values $p=\frac{2\pi\lambda}{N}$ for $\lambda\in \mathbb{Z}$ so that the Hamiltonian (\ref{xxhamilt}) takes the following form:
\begin{eqnarray}
         \tilde{H}_{XX} &=& -\frac{1}{N}\sum_{p,\,p^\prime}\,\sum_{j,k=1}^{N}V(|(j+2)-j|)u(|j-k|)e^{i(p^\prime j-pk)}c_p^\dagger c_{p^{\prime}} + H.C.~,\quad \notag \\
         H_{XX} &=& -\frac{1}{N}\sum_{p,\,p^\prime}\,\sum_{j,k=1}^{N}u(|j-k|)e^{i(p^\prime j-pk)}c_p^\dagger c_{p^{\prime}} + H.C.~. 
\end{eqnarray}
Rewriting it, we have:
\begin{eqnarray}
    \tilde{H}_{XX} = -\sum_{p}\mathcal{V}\,\tilde{u}(p)c_p^\dagger c_{p}+ H.C.~,\quad \quad  H_{XX} = -\sum_{p}\tilde{u}(p)c_p^\dagger c_{p}+ H.C.~,\label{fourierHamilt}
\end{eqnarray}
where $\mathcal{V}=V(|(j+2)-j|)$, and $\tilde{u}(p)=\sum_{l}u(l)e^{ipl}$ with $l=j-k$. Now if $j$ and $k$ are the NN indices, then $\tilde{u}(p)= \mathcal{U}e^{ip}$ with $\mathcal{U}=u(|(j+1)-j|)$. Using the results of Ref.~\cite{forbes2023exploring, Horner:2022sei, Daniel2024boy}, the complete Hamiltonians (\ref{Hamilt}) and (\ref{redHamilt}) in diagonalized form in the mean field approximation~\cite{Horner:2022sei} is written as
\begin{equation}
  \tilde{H}_{MF} = \sum_{p} \tilde{E}(p)c_p^\dagger c_p~,  \quad H_{MF} = \sum_{p} E(p)c_p^\dagger c_p~, 
\end{equation}
with $\tilde{E}(p)=\mathcal{V}E(p)$, and $E(p) = -2\,\mathcal{U} \cos (p)+ v\sin (2p)$.

\section{The continuum limit and emergent black hole background}
\label{EBHappendix}


To investigate the geometric aspects of the model and its connection to general relativity, it is essential to derive the corresponding continuum limit. This is accomplished by performing a Taylor expansion of the mean-field Hamiltonian around the Fermi points, in the vicinity of which the Dirac cone is located. Near the Fermi points, the linear part of the dispersion relation dominates, while the non-linear part becomes negligible at high momenta. As shown in Refs.~\cite{Horner:2022sei,forbes2023exploring,Daniel2024boy}, we assign lattice sites alternately to sublattices $A$ and $B$, forming a two-site unit cell, to facilitate the continuum limit analysis. This sublattice arrangement allows Dirac-like quasiparticles to emerge in the continuum limit, exhibiting a linear dispersion relation characteristic of relativistic dynamics.


The complete mean-field Hamiltonian in terms of Jordan-Wigner fermions of sub-lattices $A$ and $B$, and using Eq. (\ref{xxhamilt}) is  $\tilde{H}_{MF} = \mathcal{V}H_{MF}$, where:
\begin{equation}
  H_{MF}=-\sum_{j=1}^{N_c}\left(\mathcal{U}\left(a_j^\dagger b_j + a_j b_{j-1}^\dagger\right)+\frac{iv}{2}\left(a_j^\dagger a_{j+1} + b_j^\dagger b_{j+1}\right)\right)+ H.C.~,
\end{equation}
the fermionic operators satisfy anti-commutation relations $\{a_j,a_k^\dagger\}=\{b_j,b_k^\dagger\}=\delta_{jk}$ and all other anti-commutators vanishes, and $N_c = N/2$ is the number of unit cells. The Fourier transformation of the unit cell operators $a_j$ and $b_j$ are defined by
\begin{equation}
    a_j = \frac{1}{\sqrt{N_c}}\sum_{p\in [-\pi,\pi)}e^{ipa_c j}a_p~, \quad b_j = \frac{1}{\sqrt{N_c}}\sum_{p\in [-\pi,\pi)}e^{ipa_c j}b_p~,
\end{equation}
where, $a_c = 2a$ is the unit cell spacing for a given lattice spacing $a$. Applying this we can rewrite the mean-field Hamiltonian as 
\begin{equation}
    \tilde{H}_{MF}=\sum_{p}\left( F(p)a_p^\dagger b_p + F^\ast(p)b_p^\dagger a_p\right)+\sum_{p}G(p)\left( a_p^\dagger a_p + b_p^\dagger b_p\right)~,
\end{equation}
where $F(p)=-\mathcal{U}(1+e^{-ia_cp})\mathcal{V}$ and $G(p)=v \mathcal{V} \sin{(a_c p)}$. In terms of two-component spinor $\chi_p=(a_p, b_p)^T$ and the single particle Hamiltonian $h(p)$, the complete mean-field Hamiltonian
\begin{equation}
   \tilde{H}_{MF}=\sum_{p} \,\chi_p^\dagger h(p) \chi_p~, \quad h(p)= \begin{pmatrix}
G(p) && F(p) \\
F^\ast(p) && G(p)
\end{pmatrix}~,
\end{equation}
with the dispersion relation taking the form 
$$ \tilde{E}(p)= \mathcal{V}\left(\pm\, \mathcal{U}\sqrt{2+2\cos (a_cp)}+ v \sin (a_c p)\right).$$ 
%

{The continuum limit is derived by performing a Taylor expansion of the single-particle Hamiltonian $h(p)$ around the Fermi points. These Fermi points are defined as the momenta where the dispersion relation $\tilde{E}(p)=0$ holds, as discussed in Refs. \cite{forbes2023exploring, Horner:2022sei}. Specifically, the Fermi points are located at $p_0=\frac{\pi}{a_c}$, and $p_1=\frac{1}{a_c}\arccos{(1-\frac{2\,\mathcal{U}^2}{v^2})}$, with the second Fermi point $p_1$ existing only when the condition $|v|>|\mathcal{U}|$ is satisfied. To obtain the continuum theory, we perform a Taylor expansion of the functions $F(p)$ and $G(p)$ around the Fermi point $p_0$, retaining terms up to first order in momentum, giving}
\begin{equation}
    h(p_0+p)= a_c \,\mathcal{U}\, \mathcal{V}\sigma^y p- a_c v\mathcal{V} p\,\mathbb{I}+\mathcal{O}(p^2) \equiv e_a^l\alpha^ap_l~,
\end{equation}
where the coefficients are defined as $e_0^x=-a_c v \mathcal{V}$, $e_1^x=a_c\, \mathcal{U}\,\mathcal{V}$; and the Dirac matrices $\alpha^0=\mathbb{I}$, $\alpha^1=\sigma^y$. The continuum limit of the model is obtained by letting the unit cell spacing $a_c\rightarrow 0$ and the thermodynamic limit $N_c \rightarrow \infty$ in such a way that $N_c a_c$ is constant. Then we renormalize the couplings $a_c\, \mathcal{U} \rightarrow  \mathcal{U}$ and $a_c v \rightarrow v$, and defining the continuum limit coordinate $x=j a_c$, the continuum limit Hamiltonian after an inverse Fourier transformation to real space takes the form
\begin{equation}
    \tilde{H}=\int_\mathbb{R} dx \chi^\dagger(x)\left(-ie_a^l \alpha^a \overset{\leftrightarrow}{\partial_l}\right)\chi(x)~,\label{continuumHamilt}
\end{equation}
where we have defined the two-component spinor field $\chi(x)=\left(a(x),b(x)\right)$, the inhomogeneous couplings 
 $\mathcal{U}$ and $\mathcal{V}$ become space-dependent $\mathcal{U}\equiv \mathcal{U}(x)$ and $\mathcal{V}\equiv \mathcal{V}(x)$, $A\overset{\leftrightarrow}{\partial_\mu}B=\frac{1}{2}\left(A\partial_\mu B-(\partial_\mu A)B\right)$, and the Dirac matrices $\alpha^a=\left(\mathbb{I},\sigma^y\right)$ and $\beta=\sigma^z$. 
The corresponding action for the Hamiltonian (\ref{continuumHamilt}) is given by
\begin{equation}
    S=\int_\mathbb{M} d^{1+1}x \, \chi^\dagger(x)\left(i\overset{\leftrightarrow}{\partial_t}+ie_a^l \alpha^a \overset{\leftrightarrow}{\partial_l}\right)\chi(x)= \int_\mathbb{M} d^{1+1}x \, \overline{\chi}(x)ie_a^\mu \gamma^a \overset{\leftrightarrow}{\partial_\mu}\chi(x)~, \label{action}
\end{equation}
with $\mathbb{M}$ being the Minkowski space, $\gamma^{0}=\beta=\sigma^z$, and $\gamma^{l}=\beta\alpha^l$ implies $\gamma^{1}=-i\sigma^x$ and $ \overline{\chi}=\chi^\dagger\sigma^z$. The gamma matrices obey the anti-commutation relation $\{\gamma^{a}, \gamma^{b}\}=\eta^{ab}$ with $\eta^{ab}=\text{diag}(1,-1)$. This action closely resembles that of a Dirac field in a (1+1)-dimensional curved spacetime \cite{Morsink:1991qr, Pedernales, Yangrun}, but with two notable distinctions. First, the integration measure lacks the factor of $|e|$, effectively reducing it to the flat spacetime volume element. Second, the fields satisfy flat spacetime anti-commutation relations, given by $\{\chi_\alpha(x),\, \chi_\beta(y)\}=\delta_{\alpha\beta}\delta(x-y)$ \cite{Daniel2024boy}. To interpret this theory within the framework of curved spacetime, we introduce a spinor field $\psi = \chi/\sqrt{|e|}$ which obey the anti-commutation relation $\{\psi_\alpha(x), \,\psi_\beta(y)\}=\frac{\delta_{\alpha\beta}\delta(x-y)}{|e|}$ and propagate on a spacetime with tetrad $e_a^\mu$ with $|e|$ being the determinant of the tetrad $e_a^\mu$, which correspond to the metric $g^{\mu\nu}=e_a^\mu e_b^\nu \eta^{ab}$, explicitly given by
\begin{equation}
    e_a^\mu= \begin{pmatrix}
    1&&-v\mathcal{V}(x)\\
    0&&\mathcal{U}(x)\mathcal{V}(x)
    \end{pmatrix}~, \quad g^{\mu\nu}=\begin{pmatrix}
        1&&-v\mathcal{V}(x)\\
        -v\mathcal{V}(x)&&\mathcal{V}^2(x)\left(v^2-\mathcal{U}^2(x)\right)
    \end{pmatrix}~.
\end{equation}
When re-expressed in terms of the field $\psi$, the action of Eq. (\ref{action}) exactly corresponds to the Dirac action for a spinor field on a spacetime with the following line-element:
\begin{equation}
     ds^2 = g(x) dt^2- 2 \sqrt{1 - g(x)} \frac{dt \, dx}{\mathcal{U}(x)} -\frac{dx^2}{\mathcal{U}^2(x)}~,\label{metric}
\end{equation}
where $g(x) = 1- {v^2}/{\mathcal{U}^2(x)}$ and we have redefined the spatial coordinate $x\rightarrow x\,\mathcal{V}(x)$ since $\mathcal{V}(x)$ only acts as a conformal factor to the spatial part of the metric. 
This is the Gullstrand-Painl\'eve coordinates of spherically symmetric space-time~\cite{Shankaranarayanan:2003ya}. Employing the coordinate transformation $(t,x)\rightarrow$ $(\tau,x)\rightarrow$ $(\tau,X)$ via
\begin{equation}
    \tau(t,x)=t-\int_{x_0}^x \frac{dz}{\mathcal{U}^2(z)} \frac{v}{g(z)}~,\quad ~dX=\frac{1}{g(x)} \frac{dx}{\mathcal{U}(x)}~,
\end{equation}
brings the metric to the following two generic forms:
\begin{equation}
  ds^2 = g(x)d\tau^2 - d\mathcal{X}^2/g(x) = g(x)\left(d\tau^2 - dX^2\right),
\end{equation}
where $d\mathcal{X} = {dx}/{\mathcal{U}(x)}$ and $\mathcal{U}(x)$ is still arbitrary. 

It is important to note that the usual scaling dimension of a Dirac field $\Delta=1/2$ in a 1+1-dimensional curved spacetime can be recovered by considering the above redefinition of the spinor field $\psi = \chi/\sqrt{|e|}$. Further, it is easy to verify the scale invariance of the Hamiltonian \eqref{continuumHamilt}. Under a scaling transformation $x\rightarrow \Lambda x$, the field and couplings scale as
 \begin{equation}
     \chi(x) \rightarrow \Lambda^{-\Delta}\chi(\Lambda x)~,\quad \quad v\rightarrow \Lambda^{\beta_v} v~,\quad \quad\mathcal{U}(x)\rightarrow \Lambda^{\beta_{\mathcal{U}}}\,\mathcal{U}(\Lambda x)~, \quad \quad \mathcal{V}(x)\rightarrow \Lambda^{\beta_{\mathcal{V}}}\,\mathcal{V}(\Lambda x)~,
 \end{equation}
 $\Delta$ being the scaling dimension of the field, $\beta_v$ is the scaling dimension of the coupling $v$, and $\beta_{\mathcal{U}}$ and $\beta_{\mathcal{V}}$ are the scaling dimension of the space-dependent couplings $\mathcal{U}(x)$ and $\mathcal{V}(x)$ respectively, then the Hamiltonian 
\begin{equation}
    \tilde{H} \rightarrow \Lambda^{1-2\Delta-1} \int_\mathbb{R} dx \chi^\dagger(\Lambda x)\left(-ie_0^x \alpha^0 \Lambda^{\beta_{\mathcal{V}}+\beta_{v}+1}-ie_1^x \alpha^1 \Lambda^{\beta_{\mathcal{U}}+\beta_{\mathcal{V}}+1}\right)\overset{\leftrightarrow}{\partial_x}\chi(\Lambda x)~.
\end{equation}
For the Hamiltonian to be scale-invariant: $-2\Delta+\beta_{\mathcal{V}}+\beta_{v}+1=0$ and $-2\Delta+\beta_{\mathcal{U}}+\beta_{\mathcal{V}}+1=0$, which gives the scaling dimension of the field $\Delta=\left(\beta_{\mathcal{U}}+\beta_{\mathcal{V}}+1\right)/2$ imposing the constraint $\beta_{\mathcal{U}}=\beta_{v}$.

\subsection{Interesting limits of the BH metric}

Let us now explore the interesting limits of the metric Eq.~(\ref{metric}).

\subsubsection{Carrollian limit}

Setting $v = 0$ in the above metric, the dispersion relation near the Fermi point $p_0$ becomes $E(p+p_0)= \pm \,\mathcal{U}(x) \sqrt{(\pi -a_c p)^2}$. 
In the limit of $\mathcal{U}(x)\rightarrow 0$, the momentum-space dispersion relation flattens completely, corresponding to the Carrollian limit ($c\rightarrow 0$) \cite{Bagchi:2022eui, Banerjee:2022ocj}. This limit is associated with the collapse of the position-space light cone, where $c$ represents the speed of light.  When $v = 0$, the Hamiltonian (Eq. \ref{continuumHamilt}) simplifies to:
\begin{equation}
     H=\int_\mathbb{R} dx  \chi^\dagger(x)\left(-i\,\mathcal{U}(x) \sigma^y  \overset{\leftrightarrow}{\partial_x}\right)\chi(x)~.
\end{equation}
Up to a negative sign and the presence of $\mathcal{U}(x)$, this Hamiltonian is mathematically equivalent to a Carrollian Hamiltonian with \emph{upper gamma matrices}~\cite{Bagchi:2022eui, Banerjee:2022ocj}.  This highlights the direct connection to the Carrollian framework.  Note that $x$ is an arbitrary spatial coordinate.  Choosing a different spatial coordinate, such as $y$, would not fundamentally change the physics; the form of the Pauli matrix ($\sigma$ in this case) would simply reflect the chosen coordinate system.

\subsubsection{Conformally diverging geometry} 

In the limit of $\mathcal{U}(x)\rightarrow 0$ but with finite $v$, the metric (\ref{metric}) reduces to 
\begin{equation}
    ds^2 \approx -\frac{1}{\mathcal{U}^2(x)}\left(v^2dt^2+2v\,dtdx+dx^2\right)~=~-\frac{1}{\mathcal{U}^2(x)}\left(d\tilde{t}+dx\right)^2~,
\end{equation}
 with $\tilde{t}=v t$, and $-\frac{1}{\mathcal{U}^2(x)}$ serves as the conformal factor which diverges, making the spacetime structure poorly behaved.
 
\subsubsection{Shock wave geometry} 

If we choose $\mathcal{U}(x)$ to be  $\ln{|x-x_0|}$, $\exp{\left(\frac{1}{x-x_0}\right)}$, or similar forms, then $\mathcal{U}(x)\rightarrow\infty$ as $x\rightarrow x_0$. In such cases, the metric becomes purely timelike at $x=x_0$ for trajectories where $dt\neq 0$. This situation can be associated with a gravitational shock wave, as these functions introduce discontinuities in the metric similar to those produced by gravitational shock waves generated by massless \cite{Dray:1984ha, Srivastava:2020cdg}, or massive particles \cite{Mackewicz:2021jfr}. However, unlike typical gravitational shock waves, which are associated with null hypersurfaces, the shock wave in this case is associated with timelike hypersurfaces. This distinction is crucial: timelike hypersurfaces permit two-way communication across the boundary, while null hypersurfaces are causally one-directional, corresponding to signals propagating at the speed of light.

The discontinuity in the metric could also represent phenomena such as a domain wall or an interface between two regions of spacetime with different physical properties.  It might also correspond to a sharp concentration of matter or energy localized on a timelike hypersurface, such as a thin shell. 

However, in the main text we considered $\mathcal{U}(x)$ to be free of such discontinuities, ensuring a well-behaved metric throughout the region of interest. In particular, $\mathcal{U}(x) \sim \sech(x)$ is particularly suitable due to its smoothness and exponential decay, avoiding singularities and providing a consistent and physically meaningful framework.


\section{BH critical exponents}

Near the threshold of BH formation, a striking phenomenon known as critical behavior emerges~\cite{choptuik1993universality}. Much like in phase transitions, the properties of a forming BH do not evolve smoothly but instead follow power-law scaling. As an initial condition parameter ($\eta$) 
--- such as the strength of an incoming matter wave --- approaches a critical value ($\eta_*$), the resulting BH’s mass ($M_{BH}\propto (\eta - \eta_{\ast})^{\gamma_M}$ where ${\gamma_M}$ is the critical exponent) follows a power-law dependence, characterized by a critical exponent~\cite{garfinkle,gundlach}. Remarkably, these exponents, along with others governing different physical quantities, often exhibit \emph{universality}, meaning they remain the same regardless of the specific nature of the collapsing matter. This universality suggests a deeper underlying structure within Einstein’s equations and points to intriguing connections between BH formation and broader studies of critical phenomena, potentially shedding light on the quantum aspects of gravity~\cite{welsh2018simple, christo1,sorkin}.


Broadly there are two regimes in this scenarios: First, the supercritical regime, where black holes of arbitrarily small mass can form. In this regime, the BH mass $M_{BH}$ scales as $M_{BH}\propto (\eta - \eta_{\ast})^{\gamma_M}$ where $|\eta - \eta_{\ast}|^{\gamma_M}$ has the dimensions of length. Second, the subcritical regime, where the maximum scalar curvature $R_{max}$ before the field disperses scales as $R_{max}\propto (\eta - \eta_{\ast})^{-2\gamma}$~\cite{garfinkle,gundlach}. 
If our analog model truly mimics black hole formation, it should reproduce these critical exponents. Therefore, it is of interest to explore the relationship between the critical exponents $\gamma_M$ and $\gamma$ within the context of the analog metric (\ref{metrictau}), specifically in $D=2$. While the relation $\gamma_M=(D-3)\gamma$ has been established for higher dimensions ($D\geq 4$) \cite{sorkin}, we investigate whether a similar or modified scaling relation applies in $D = 2$.

\begin{figure}
    \centering
    \includegraphics[width=0.3\linewidth]{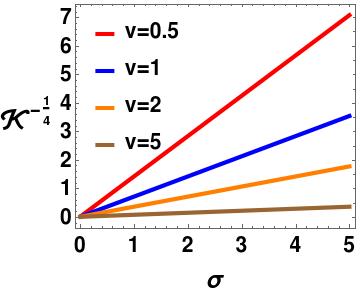}
    \caption{Inverse fourth root of the Kretschmann scalar, $\mathcal{K}^{-1/4}=\sigma\sqrt{\sigma_c}$ as a function of $\sigma$ for four different values of $v$ = 0.5, 1, 2, and 5. }
    \label{Kretschfig}
\end{figure}

\subsection{Kretschmann scalar and surface gravity}
\label{Kretschmann exponent}

In our study, we focus on the Kretschmann scalar ($\mathcal{K}$), a quadratic scalar invariant having dimensions $[L]^{-4}$. Defined as $\mathcal{K} = R_{\mu\alpha\beta\rho}R^{\mu\alpha\beta\rho}$, the Kretschmann scalar, unlike the Ricci scalar, remains non-zero in vacuum regions, providing valuable information where the Ricci scalar is trivial. For the metric (\ref{metrictau}), we find the Kretschmann scalar, $\mathcal{K}=1/(\sigma^4 \sigma_c^2)$ which is independent of $x$.

Since $\mathcal{K}$ is constant in this scenario, a standard scaling analysis is not possible, and thus $\mathcal{K}$ cannot directly yield critical exponents. However, as shown in Fig. \ref{Kretschfig}, our numerical results confirm a linear relationship between $\mathcal{K}^{-\frac{1}{4}}$ 
and $\sigma$ for different values of $v$. This linearity suggests a connection to the critical behavior, as  $\sigma-\sigma_c$ quantifies the proximity to the critical point.  The absence of scaling in $\mathcal{K}$ indicates that it is likely not the optimal quantity for directly determining the correlation length exponent.

This leads us to surface gravity, which is defined as
\begin{equation}
    \kappa_g^2 = -\frac{1}{2}(\nabla_\mu K_\nu)(\nabla^\mu K^\nu)~,
\end{equation}
$K^\mu$ is the Killing vector associated with the Killing horizon $\Sigma$, and $n^\mu$ is the unit normal vector to the horizon surface.
According to the zeroth law of BH mechanics, the surface gravity of a stationary BH is constant over the event horizon and independent of the observer. To confirm the scaling exponent of the surface gravity, we evaluate it at the horizon. For the metric (\ref{metrictau}), the components of the Killing vector are: $ K^\mu = (1,0)$ and $K_\mu = \left(g(x),0\right)$, while the components of the the normal vector are:
\begin{equation}
    n^\mu = \left(\sqrt{g(x)},0\right)~, \qquad n_\mu = \left(\frac{1}{\sqrt{g(x)}},0\right)~.\label{normalvec}
\end{equation}
These vectors are normalized such that $K^\mu K_\mu=n^\mu n_\mu=1$. At the event horizon $(v=\mathcal{U})$, we can verify that $K^\mu K_\mu=0$, $n^\mu n_\mu=0$, and $K^\mu n_\mu=0$. Thus, the Killing vector is not only null but also orthogonal to the horizon itself, confirming that the event horizon is a Killing horizon. The surface gravity is then given by:
\begin{equation}
    \kappa_g^2 = \frac{1}{8 \sigma \, \sigma_c^2} \left[6
\sinh ^2\left(\frac{x}{\sigma }\right) - 
  \frac{1}{\sigma} \tanh ^2\left(\frac{x}{\sigma }\right) g^2(x) \right] \,~.
\end{equation}
Using this expression of $\kappa_g^2$, we analyze its critical exponent and compare it with the correlation length exponent  $\nu_P$ computed in subsection \ref{percoexpo}. 
The scaling behavior is numerically validated and presented in Fig. \ref{CorrelateKretsch} (right), where we plot $\log \kappa_g^2$ as a function of $\log |x-x_c|$ Here, $x$ is chosen close to $x_c$ for specific values of $\sigma$ and $v$, corresponding to spacetimes near the threshold of BH formation (subcritical collapse). In Fig. \ref{CorrelateKretsch} (right), the filled circles are the analytical values obtained through Eq. (\ref{surfg}), while the solid lines are the numerical fits confirming the linear scaling behavior. For the red curve, corresponding to parameter values $v=0.1$ and $\sigma=50.1$, the linear fit follows the relation $\kappa_g^2=(x-x_c)^{2.00}$. Similarly, for the blue curve, with $v=0.5$ and $\sigma=2.1$, the linear fit also adheres to $\kappa_g^2=(x-x_c)^{2.00}$, up to an unimportant constant in each case. These results demonstrate that the critical exponent associated with the gravitational field strength, represented by the surface gravity, is $\gamma=1.00$. 

Interestingly, we observe that the correlation length exponent $\nu_P=1.11$ for the NNN dominating phase, referred to as the \emph{black hole phase} in our percolation model, closely matches the exponent $\gamma=1.00$ associated with the gravitational field strength. This agreement suggests that the formation of clusters in the percolation model mimics the structure and dynamics of the gravitational collapse leading to BH formation.


\subsection{Komar mass}\label{Komar exponent}

To make the connections between the percolation model and gravitational collapse more concrete, we now turn to the Komar mass, which represents the total mass (or energy) contained within the spacetime, as seen from infinity. The Komar mass $M_{K}$ for $(1 + 1)-$D space-time in terms of Killing vector $K^\nu$ is given by~\cite{carroll2019spacetime},
\begin{equation}
    M_{K}=\int_S dx \sqrt{|g|} n_\mu R^{\mu\nu} K_\nu~,
\end{equation}
where $S$ is a $1$-dimensional constant time slice, $g$ is the determinant of the  $1$-dimensional metric of Eq. (\ref{metrictau}), $n^\mu$ is a unit vector orthogonal to $S$, and $R^{\mu\nu}$ is the Ricci tensor. Using the Killing equation: $\nabla^\mu K^\nu + \nabla^\nu K^\mu = 0$ and the Killing vector lemma: $\nabla_\nu \nabla_\mu K^\mu = R_{\mu\nu} K^\nu$, the Komar mass can be expressed as
\begin{equation}
    M_{K}=\int_S dx \sqrt{|g|} n_\mu \nabla_\nu(-\nabla^\nu K^\mu)~,
\end{equation}
with the forms of $K^\mu$ and $n_\mu$ given in Eq. (\ref{normalvec}). The boundary of $S$ has two contributions, one from an inner boundary at the horizon which is the mass of the BH, denoted by $M_{BH}$ and the second from an outer boundary at infinity which is the contribution of matter outside the event horizon to the total mass, denoted by $M_{\infty}$.  With these inputs, we can write the Komar mass: $M_K=M_{BH}+M_{\infty}$, where
\begin{eqnarray}
    M_{BH}&=& \frac{1}{\sigma_c} \sqrt{\frac{2}{\sigma }}  \tan ^{-1}\left[\tanh \left(\frac{1}{2} \cosh
   ^{-1}\left[\sqrt{\frac{\sigma_c}{{\sigma }}} \right]\right)\right],\\
   M_{\infty}&=&-\frac{1}{\sqrt{2} \sigma_c\,\sigma ^{3/2}}\int_{S\rightarrow\infty}dx\,\, \text{sech}\left(\frac{x}{\sigma }\right)~.
\end{eqnarray}

We analyze the numerical scaling of the BH mass ($M_{BH}$) with $\sigma$, which determines proximity to the critical point ($\sigma_c$).  Since $\sigma - \sigma_c$ quantifies the distance from this threshold, this scaling reveals insights into the critical behavior of the system and gravitational collapse dynamics. Figure \ref{EnergyKomar} (right) shows analytical values (circles) and numerical fits (lines) validating the scaling. For $v=0.1$ (red) and $v=0.5$ (blue), the fits follow $M_{BH} \propto (\sigma-\sigma_c)^{0.5}$, confirming a critical exponent $\gamma_M \approx 0.5$. This is in close agreement with  $\gamma_M = 0.53$ found in 2-D dilaton gravity~\cite{peleg1997}.

Two key observations emerge from our analysis. First, the critical exponent $\beta = 0.53$ for total energy in the NNN-dominant phase (see subsection \ref{percoexpo}) 
aligns with $\gamma_M$, suggesting that clustering dynamics in the percolation model closely resemble the structural and dynamical features of gravitational collapse leading to BH formation, highlighting the relevance of percolation theory in capturing key aspects of gravitational field scaling. Table \ref{tab:critical_exponents} compares these exponents.
Second, the relationship between the critical exponents $\gamma_M$ and $\gamma$ for our BH metric Eq. (\ref{metrictau}) in $D=2$ spacetime dimensions deviates from the conventional scaling relation $\gamma_M=(D-3)\gamma$ outlined in \cite{sorkin} for $D\geq 4$, adopting a modified form $\gamma_M=2\gamma$. 
This requires further investigation. 


\section{Hoop conjecture}

Thorne's hoop conjecture~\cite{thorne2000gravitation} postulates that a mass $\mathcal{M}$ collapses to form a BH if and only if it can be enclosed by a circular hoop of critical circumference $C_{critical} = 4\pi\mathcal{M}$~\cite{flanagan1991hoop,hod2018status,bonnor1983hoop}.  This implies $C \leq 4\pi\mathcal{M}$ for BH formation.


We explore this conjecture within the percolation model, associating the the length of the smallest percolation cluster with the circumference $C$, and the energy of the smallest cluster in the percolation model will be identified with the mass $\mathcal{M}$ specified by the hoop conjecture. The length of the smallest cluster is defined as the number of distinct states within the cluster, while the energy of the smallest cluster corresponds to the number of connected bonds it contains. 

Figure \ref{Hoopfig} is the plot of the number of connected bonds, $E_b$, and the number of states $C$ in the smallest cluster for both  NN (red) and NNN (blue) dominated phases. The red and blue dots refer to the numerical output for different lattice sizes $3 \leq N \leq 20$. The results reveal that the slopes of the plotted lines for both the NN and NNN dominating phases are equal.
The equal slopes suggest $\mathcal{U} \geq 1$ (NN) and $v \geq 1$ (NNN), but these do not necessarily imply $v \geq \mathcal{U}$, a BH formation requirement.  Thus, the conjecture does not favor one phase over the other.

While the hoop conjecture holds in both phases, BH formation occurs only in the NNN phase. This is due to the exponential growth of both cluster number and energy in the NNN phase, facilitating gravitational collapse.  The NN phase exhibits only linear growth, preventing BH formation.

\bibliography{lattice}

\end{document}